\documentclass[12pt,a4paper,dvips]{article}
\usepackage{a4p}
\usepackage{cite,mcite}
\usepackage{graphicx}
\usepackage{rotating}
\usepackage{physics}
\usepackage{l3_title,ifthen,Lep}
\journalname{Phys. Lett. B}
\date{September 7, 2001}
\preprint{2001-065}
\newlength{\capindent}
\newlength{\capwidth}
\newlength{\figwidth}
\newcommand{\icaption}[2][!*!,!]{\hspace*{\capindent}%
  \begin{minipage}{\capwidth}
    \ifthenelse{\equal{#1}{!*!,!}}%
      {\caption{#2}}%
      {\caption[#1]{#2}}
  \end{minipage}}
\newcommand{\GG}{\ensuremath{\gamma ^* \gamma ^*}}
\newcommand{\Gg}{\ensuremath{\gamma \gamma}}
\newcommand{\pz}{\ensuremath{\pi^0}}
\newcommand{\ks}{\ensuremath{\rm{K_S^0}}}
\newcommand{\Wgg}{\ensuremath{\W_{\Gg}}}

\newcommand{\sqs}{\ensuremath{\sqrt{s}}}

\newcommand{\dpt} {\ensuremath{d\sigma / dp_t}}
\newcommand{\deta} {\ensuremath{d\sigma / d|\eta|}}
\begin{document}
\bibliographystyle{l3style}
\begin{titlepage}
\title{Inclusive {\boldmath \pz} and {\boldmath \ks} Production \\ 
in Two-Photon Collisions at LEP}
\author{The L3 Collaboration}
%
%
\begin{abstract}
The reactions $\rm{e}^{+} \rm{e}^{-} \rightarrow \rm{e}^{+} \rm{e}^{-} \pz $ X
and $\rm{e}^{+} \rm{e}^{-} \rightarrow \rm{e}^{+} \rm{e}^{-} \ks $ X
are studied 
using data collected at LEP with the L3 detector  
at centre-of-mass energies between 189 and 202 GeV. 
Inclusive differential cross sections
are measured as a function of the particle transverse momentum \pt {} and
the pseudo-rapidity.
For $ \pt \le 1.5 \GeV $, the \pz {} and \ks {} differential cross sections are 
described by an exponential, typical of soft hadronic processes.
For $ \pt \ge 1.5 \GeV$, the cross sections show the presence of perturbative QCD processes,
described by a power-law. The data are
compared to  Monte Carlo predictions and to NLO QCD calculations. 

\end{abstract}

\submitted 

\end{titlepage}
%
%
\section{Introduction}

Two-photon collisions are the main source of hadron production 
in the high-energy regime of LEP 
 via the process ${\rm e}^{+} {\rm e}^{-} \rightarrow {\rm e}^{+} {\rm e}^{-}
 \gamma ^{*}  \gamma ^{*}  \rightarrow 
 {\rm e}^{+} {\rm e}^{-}  hadrons$.
In the Vector Dominance Model, each photon can transform
into a vector meson with the same quantum numbers, 
thus initiating a strong interaction process with characteristics 
similar to hadron-hadron interactions. This
process dominates in the ``soft'' interaction region, where hadrons are produced
  with a low  transverse momentum \pt.
Hadrons with  high \pt {} are produced by the direct QED process $\GG \ra \qqbar$ 
or by QCD processes originating from the partonic content of the photon.
QCD calculations are available 
for single particle inclusive production in two-photon interactions at 
next-to-leading order (NLO)  \cite{gordon1,kniehl}.

\par
In this letter, inclusive \pz {} and \ks {} production from
quasi-real photons is studied for a  
centre-of-mass energy  of the two interacting photons, \Wgg ,  
greater than 5 \GeV. The $\pi^0$'s are measured 
in the transverse momentum range $0.2 \le \pt \le 20 \GeV$ and in 
the pseudo-rapidity\footnote{$\eta  =  -\ln ~ \tan (\theta / 2)$, where 
$\theta$ is the polar angle of the particle relative to the beam axis.}
interval
$|\eta| \le 4.3$.
The \ks 's are measured in the range $0.4 \le \pt \le 4 ~ \rm{GeV}$ and 
$|\eta| \le 1.5$.

\par 
The data used for this analysis were collected by the L3 detector \cite{L3}
 at centre-of-mass energies from \sqs {} = 189 \GeV {} to 202 \GeV , with a
luminosity weighted  average 
value of \sqs {} = 194 \GeV.
The integrated luminosity is 414 \pb .
Previous measurements of inclusive charged hadron and \ks {} production
were performed at LEP \cite{opal} at \sqs {} = $161-172$ \GeV .

\section{Monte Carlo simulation}
\par
The  process ${\rm e}^{+} {\rm e}^{-} \rightarrow {\rm e}^{+} {\rm e}^{-}   hadrons$  
is modelled with the PHOJET  \cite{Engel} and PYTHIA  \cite{PYTHIA}
event generators with respectively 2 and 3 times more luminosity than the data.
The following generators are used to simulate background processes: PYTHIA
and KK2f\cite{KK2f} for 
\ee $\rightarrow \rm q \bar{q} \,(\gamma $); 
KORALZ \cite{KORALZ} for \ee $\rightarrow \tau^{+} \tau^{-}(\gamma )$;
KORALW \cite{KORALW} for \ee $\rightarrow \rm{W}^{+} \rm{W}^{-}$
and  DIAG36 \cite{DIAG36} for \ee \ra {} \ee $\tau^{+} \tau^{-}$.
The events are simulated in the L3 detector using the GEANT \cite{GEANT}
and GEISHA \cite{GEISHA} programs
and passed through the same reconstruction program as the data.
Time dependent detector inefficiencies, as monitored during the data taking
period, are also simulated.

\section{Event selection}

\par
The selection of  ${\rm e}^{+} {\rm e}^{-} \rightarrow 
 {\rm e}^{+} {\rm e}^{-} hadrons$ events is based on information from 
the central tracking detectors and from 
the electromagnetic (BGO)  and hadronic   calorimeters \cite{l3tot}.
 In order to restrict the $Q^2$ interval, we exclude events with
a cluster in the small-angle calorimeter with energy greater than 30 GeV.
About 2 million  hadronic events are selected.
The level of background  is less than  1\% and is 
mainly due to the $\ee \rightarrow \rm q \bar{q} \,(\gamma)$ and
 \ee \ra {} \ee $\tau^{+} \tau^{-}$  processes.

\par
The particle identification proceeds from charged tracks and 
electromagnetic clusters. 
The inner tracking detector extends up to $|\eta| = 1.64$. 
The electromagnetic calorimeters
extend up to $|\eta| \le 0.96$ for the barrel, and cover 
$1.15 \le |\eta| \le 2.25$ for the endcaps and 
$3.37 \le |\eta| \le 4.38$ for the small-angle detector.
A  track must have 
a transverse momentum above 100 \MeV {} and a distance of closest 
approach  to the primary vertex in the transverse plane below 10 mm. 
An electromagnetic cluster must have an energy greater than 100 \MeV {} 
 formed by the energy deposited in at least 2 neighbouring
 BGO crystals. There should be no charged track 
within an angle of 200 mrad and 
the associated energy  in the hadron calorimeter must be less
than 20\% of the electromagnetic energy.
Clusters in the small-angle detector must have an energy greater than 2 GeV 
and restrictions on the energy profile
in each cluster are applied to distinguish well reconstructed photons 
from those at the edges of the
detector or from residual hadrons.

\par
\begin{figure}[htbp]
  \begin{tabular}{cc}
  \includegraphics[width=0.45\figwidth]{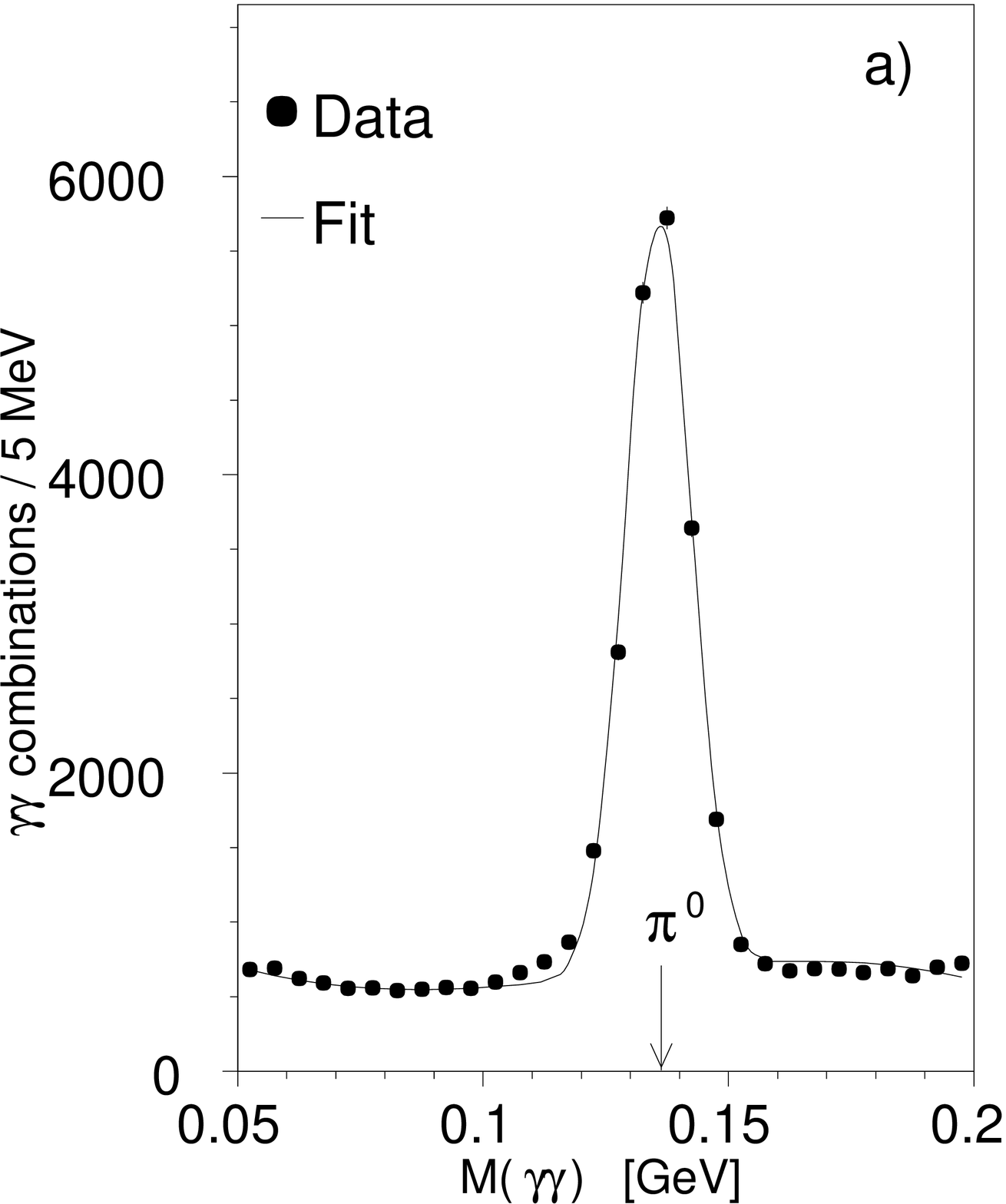}&
  \includegraphics[width=0.45\figwidth]{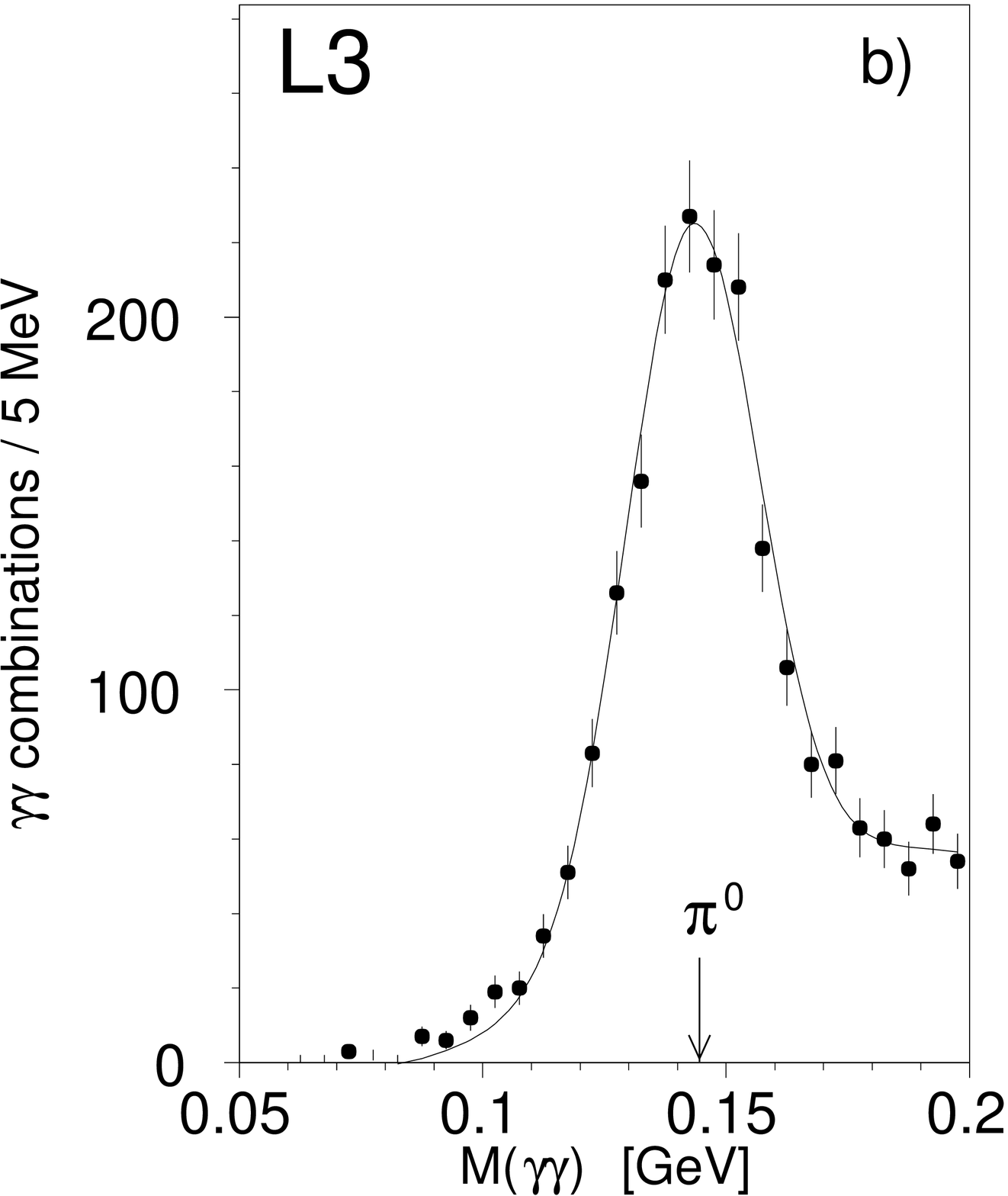}\\
\\
  \includegraphics[width=0.45\figwidth]{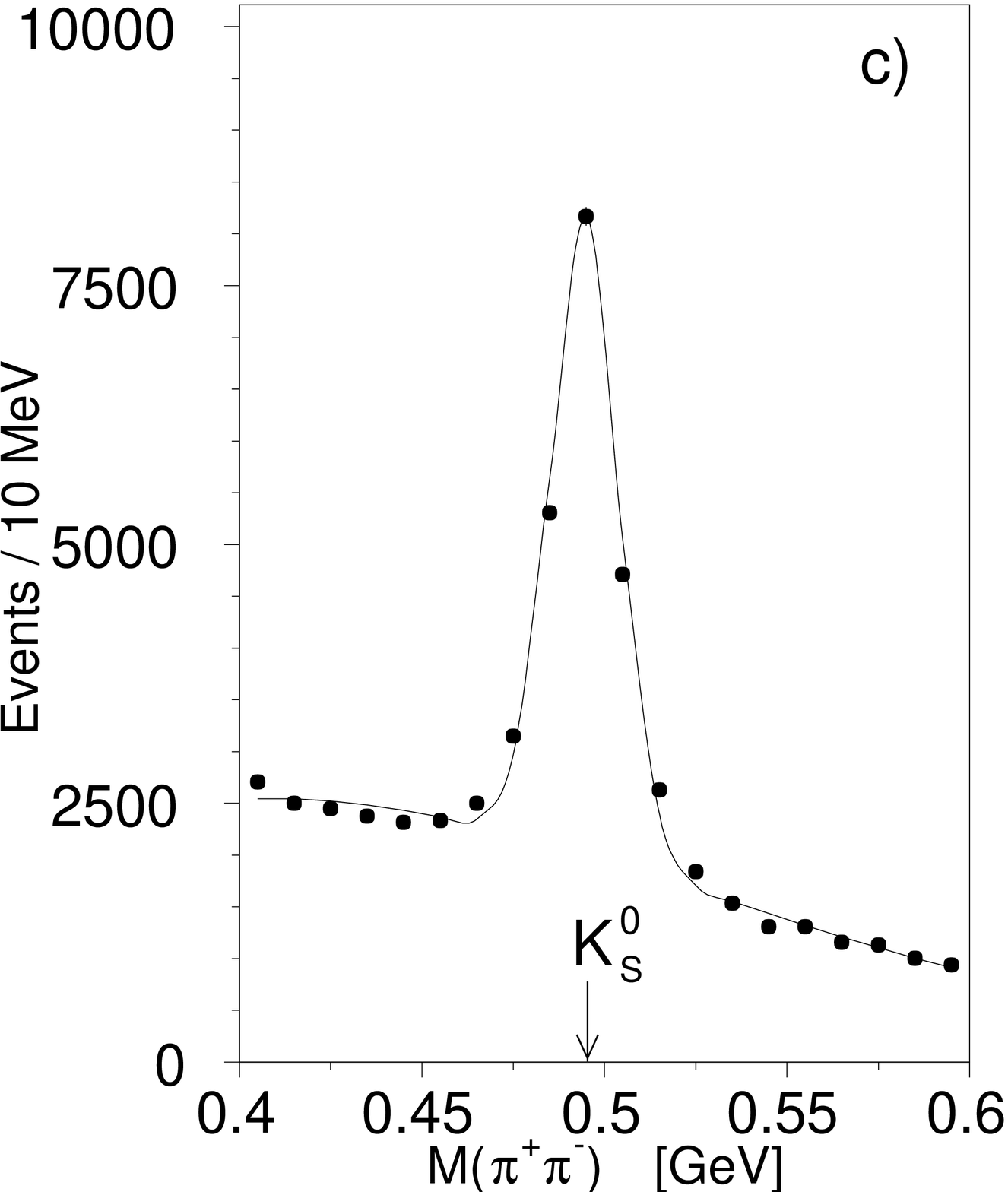}&
  \includegraphics[width=0.45\figwidth]{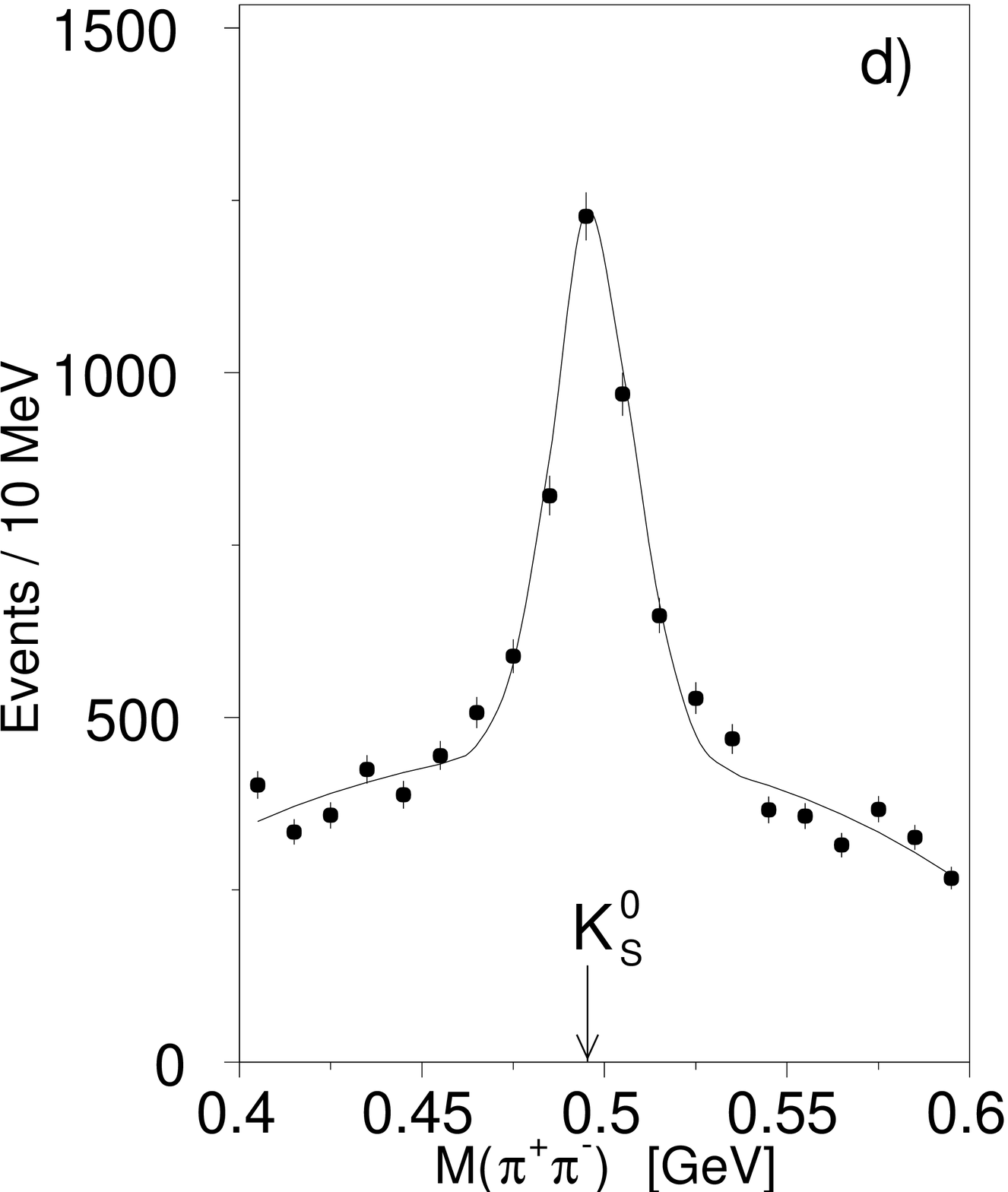}\\
  \end{tabular}
  \caption{Two  photon effective mass for a) $1 < \pt < 1.5$ GeV in the central region
  and b) for \mbox{$\pt  > 0.2$ GeV} in the small angle detector.
  Two charged pion effective mass for c) \mbox{$0.2 <  \pt < 0.4$ GeV } and
  d) $0.8 < \pt < 1.0$ \GeV.
  The \pz {} and \ks {} peaks are fitted with a Gaussian
  and the background with a Chebyshev polynomial. Values of the \pz {} and \ks {} masses
  are also indicated.}
  \label{fig:fit}
\end{figure}

For $\pt <$ 5 GeV, the
inclusive  $\pi^0$ production is measured via the decay of the 
\pz {} into two photons associated to two electromagnetic clusters.
The distribution of the effective mass of the 
reconstructed \Gg {} system 
shows a clear $\pi^0$ peak in all the detector regions.
Examples for the central region and the small-angle detector
are given in Figures \ref{fig:fit}a and \ref{fig:fit}b, respectively.
Over the entire range of $|\eta|$ and \pt, 
the resolution varies between 6.6 and 13.5 MeV,
and is well reproduced by Monte Carlo simulation.
For $\pt > 4$ \GeV {} and $|\eta|< 0.5$, 
the two final photons are unresolved and the \pz {} is associated
to a single electromagnetic cluster.
To avoid double-counting in the region $4 < \pt < 5 \GeV$ and $|\eta|< 0.5$,
where both methods are applied,
only clusters which do not contribute to combinations  
in a 3-$\sigma$ mass band around the $\pi^0$ peak are taken into account.
In this region,
we have checked that the two methods applied separately agree within errors.

\par
Inclusive \ks \, production is measured using the  decay 
 $ \ks \ra \pi^+ \, \pi^-$  that produces two oppositely charged tracks.
The \ks's are selected by reconstructing the secondary decay vertex. 
The projected distance, in the transverse plane, 
 between the secondary vertex and the primary $\ee$ interaction point 
is required to be greater than 3 mm. 
The angle between the projected flight direction of the \ks {} candidate
and the total transverse momentum vector of the
two outgoing tracks must be less than 75 mrad.
After these cuts, about $5\times 10^5$ events are selected. 
The distribution of the effective mass of the 
reconstructed $\pi^+ \, \pi^-$ system 
shows a clear \ks {} peak.
Examples for different \pt {} bins
are given in Figures \ref{fig:fit}c and \ref{fig:fit}d.
The resolution varies  from 8 MeV for \pt {} $<$ 1 GeV  to 
\mbox{10 MeV} around
4 GeV, and is well reproduced by Monte Carlo simulation.

\section{Differential cross sections}

\par 
Differential cross sections  as a function of the transverse momentum \pt {} 
and  of the absolute pseudorapidity $|\eta|$ 
are calculated using the number of \pz {} and
\ks {} candidates and the overall efficiency for each  bin  of \pt{} or $|\eta|$.
The overall efficiency includes reconstruction and trigger efficiencies and 
takes into account the branching fraction of the $\kos$
into $\pip\pim$.
The reconstruction efficiency includes the effects of the acceptance and 
selection cuts and is calculated with the Monte Carlo
generators PHOJET and PYTHIA.
As both generators reproduce well the shapes of the experimental distributions 
of hadronic two-photon production\cite{l3tot},
their average is used.

\par
Two-photon  events  are collected predominantly by the track triggers \cite{tracktrig}.
The trigger efficiency is derived from each year's  data sample  
by comparing  the number of events accepted by 
the independent track and   
calorimetric energy\cite{etrig} triggers. 
The efficiencies of higher level triggers are measured using
prescaled  events. 
For the \pz, it varies from
80\% at low \pt {} to 100\% at high \pt .
For the \ks, it is 85\% independently of \pt .

\par
The cross sections are calculated for $\Wgg \ge 5 \GeV $
and a photon virtuality $Q^2 \le 8 ~ \rm{GeV} ^2 $. 
The overall efficiency does not depend on the  $Q^2$ cutoff.

\subsection{{\boldmath $ \rm{e}^{+} \rm{e}^{-} \rightarrow \rm{e}^{+} \rm{e}^{-} \pz$}  X analysis}

\par
To evaluate the number of \pz's  
when the two photons are well separated in the detector,
fits are made to the $\Gg$ mass distribution  
in the interval $50 < M_{\Gg} < 200$  \MeV {}
using a Gaussian to describe the
signal and a third degree Chebyshev polynomial for the background. 
All the parameters, including mass and width of the peak, are left free. 

\par 
When single clusters are identified as a \pz,  
the contamination coming from the decays of
other mesons ($\eta$, $\omega$, $\eta'$,...)
is on average \mbox{$15.1\pm1.2$ \%} over the entire \pt {} and $|\eta| $ ranges.
Single photon production ($\rm \gamma q \rightarrow \gamma q$, $\rm q \bar{q} \rightarrow \gamma g$,
$\rm g q \rightarrow \gamma q$) is predicted to be 
more than one order of magnitude below our
measurements \cite{gordon2}. In addition, a study of the energy profile
of each cluster reveals no significant background from this source.
The background due to annihilation events increases with \pt {} up to 
a maximum of 11 \%.

\par
The reconstruction efficiency 
 varies between 15\% and 50\% in the different \pt {} and $|\eta|$ ranges. 
The efficiency increases from  $\pt \simeq 0.2 \GeV$, where a low energy 
photon can go undetected, up to  $\pt \simeq 2 \GeV$.  
In the region $2 < \pt < 4 \GeV$, 
the efficiency decreases due to the increasing percentage of events 
in which the two photons merge. 
For $\pt > 4 \GeV$, the addition of the single-cluster analysis gives a higher efficiency.
The efficiency decreases with polar angle due to the acceptance of the calorimeters.

\par
Sources of systematic uncertainties on the cross-section measurements
are selection criteria, Monte Carlo modelling, background subtraction 
and accuracy of the trigger efficiency  measurement.
The uncertainty due to  selection criteria is dominated by hadron selection, 
estimated  to be 7.5 \% \cite{l3tot}. 
The Monte Carlo modelling uncertainty,
taken as half the relative difference  between PHOJET and PYTHIA,
increases with \pt {} from 1\% to 24\%.
The background uncertainty varies from 5\% to 15\% for \mbox{$\pt <$ 5 GeV}.
It is estimated using different background parametrisations during the fitting
procedure.
In the high \pt {} region, the uncertainty on the annihilation background
subtraction is taken as half the difference between PYTHIA and KK2f and varies from
0.1\% to 5\%.
The uncertainty on the trigger efficiency varies  from 0.1\% to 1.1\% 
due to the statistical accuracy of its determination.

\par
The overall efficiencies and differential cross sections \dpt {} and \deta {} are given in
Tables \ref{tab:pi0pt} and \ref{tab:pi0eta}. 
The \pz {} multiplicity in the range  \mbox{$0.2 < \pt < 20$ \GeV} {} and \mbox{$|\eta| <$ 0.5}
is  \mbox{$0.275\pm0.001\pm0.025$} per $\ee \ra \ee hadrons$ event, in
agreement with Monte Carlo predictions, 0.281 for PHOJET and 0.285 for PYTHIA.

\subsection{{\boldmath $ \rm{e}^{+} \rm{e}^{-} \rightarrow \rm{e}^{+} \rm{e}^{-} \ks $} X analysis}

The number of \ks {} is evaluated by means of a fit 
to the $\pi^+ \pi^-$ mass distribution 
in the interval $400 < M_{\pi^+\pi^-} < 600$  \MeV.
A Gaussian describes the signal and a 
third degree Chebyshev polynomial the background. 
All parameters, including the mass and width of the peak, are left free. 

The reconstruction efficiency is of the order of 20 \%.
Systematic uncertainties, estimated as in the \pz {} case, 
are selection criteria  (7.5\%),
Monte Carlo modelling (1$-$6\%), 
background subtraction (1$-$7\%)
and trigger efficiency measurement accuracy (2\%).
In addition, a 2.5 \% uncertainty arises from the \ks {} selection criteria.

\par
The  overall efficiencies and differential cross sections \dpt {} 
and \deta {} are given in Tables \ref{tab:k0spt} and \ref{tab:k0seta}.
The multiplicity of \ks {} in the range  $0.4 < \pt < 4$ \GeV {} and $|\eta| <$ 1.5
is $0.060\pm0.006\pm0.003$ per \mbox{$\ee \ra \ee hadrons$} event, 
in agreement with Monte Carlo predictions,
0.067 for PHOJET and 0.056 for PYTHIA.

\section{Results}

\par
\begin{figure}[htbp] 
 \begin{tabular}{cc}
    \includegraphics[width=0.45\figwidth]{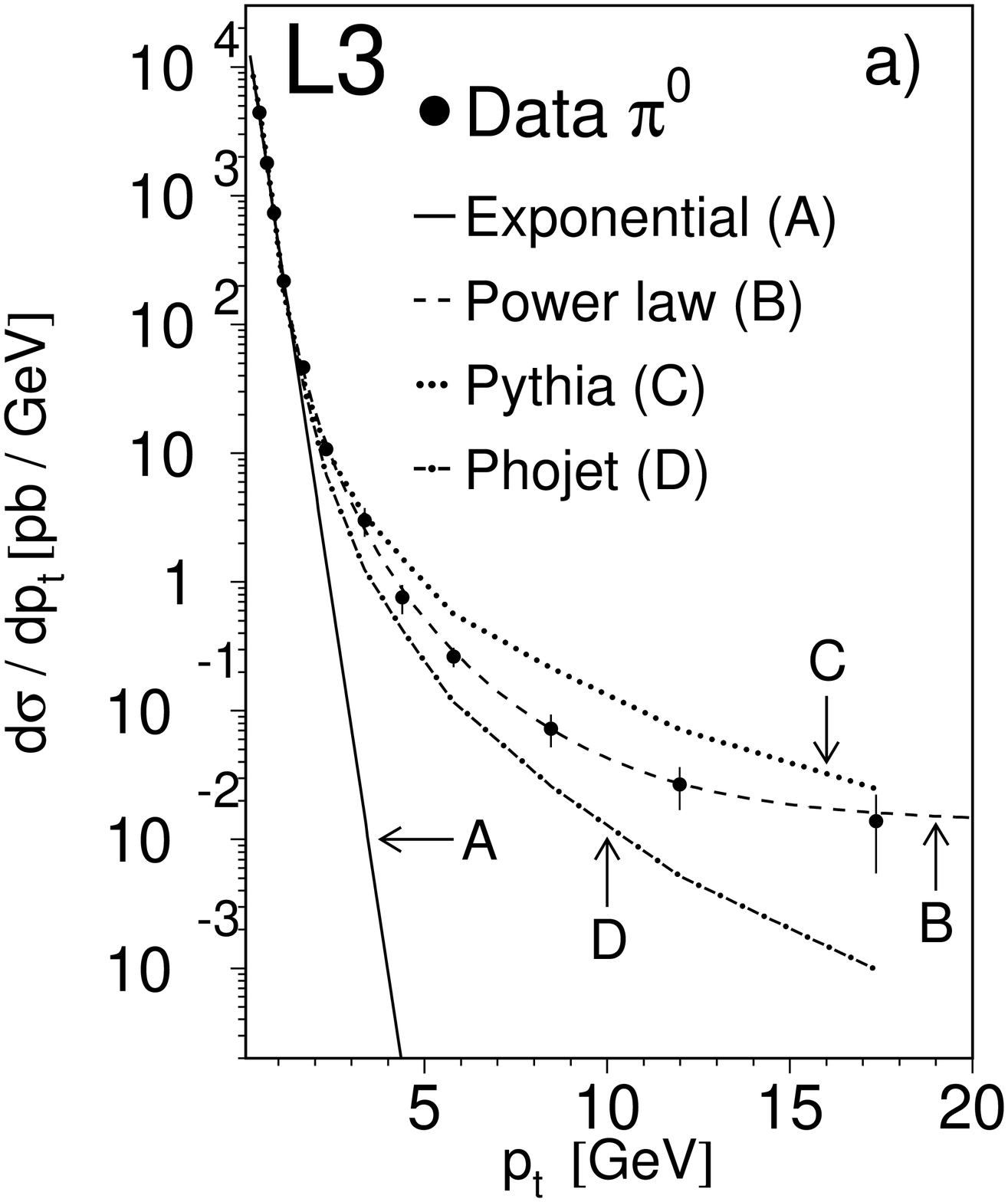}&
    \includegraphics[width=0.45\figwidth]{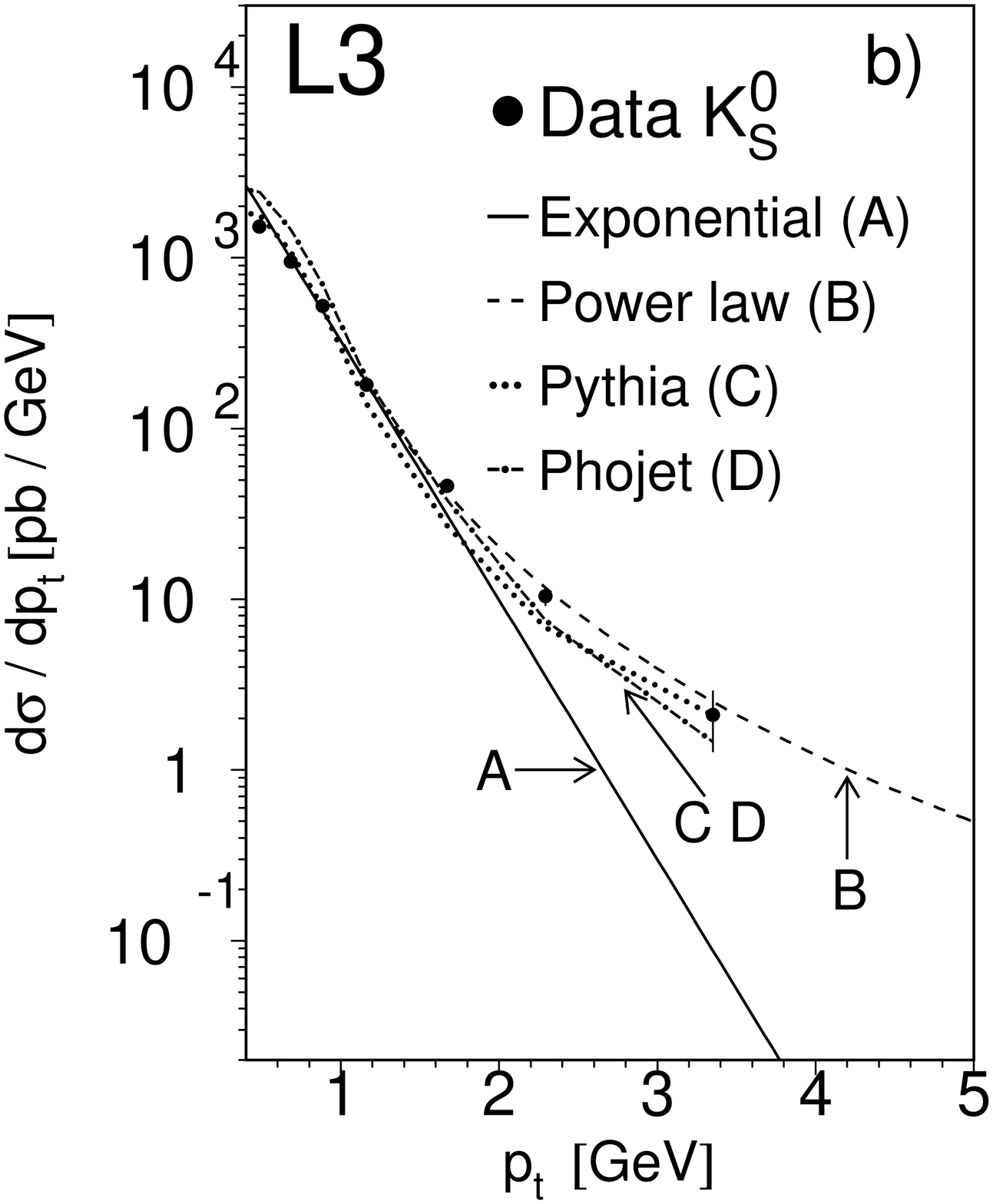}\\
    \includegraphics[width=0.45\figwidth]{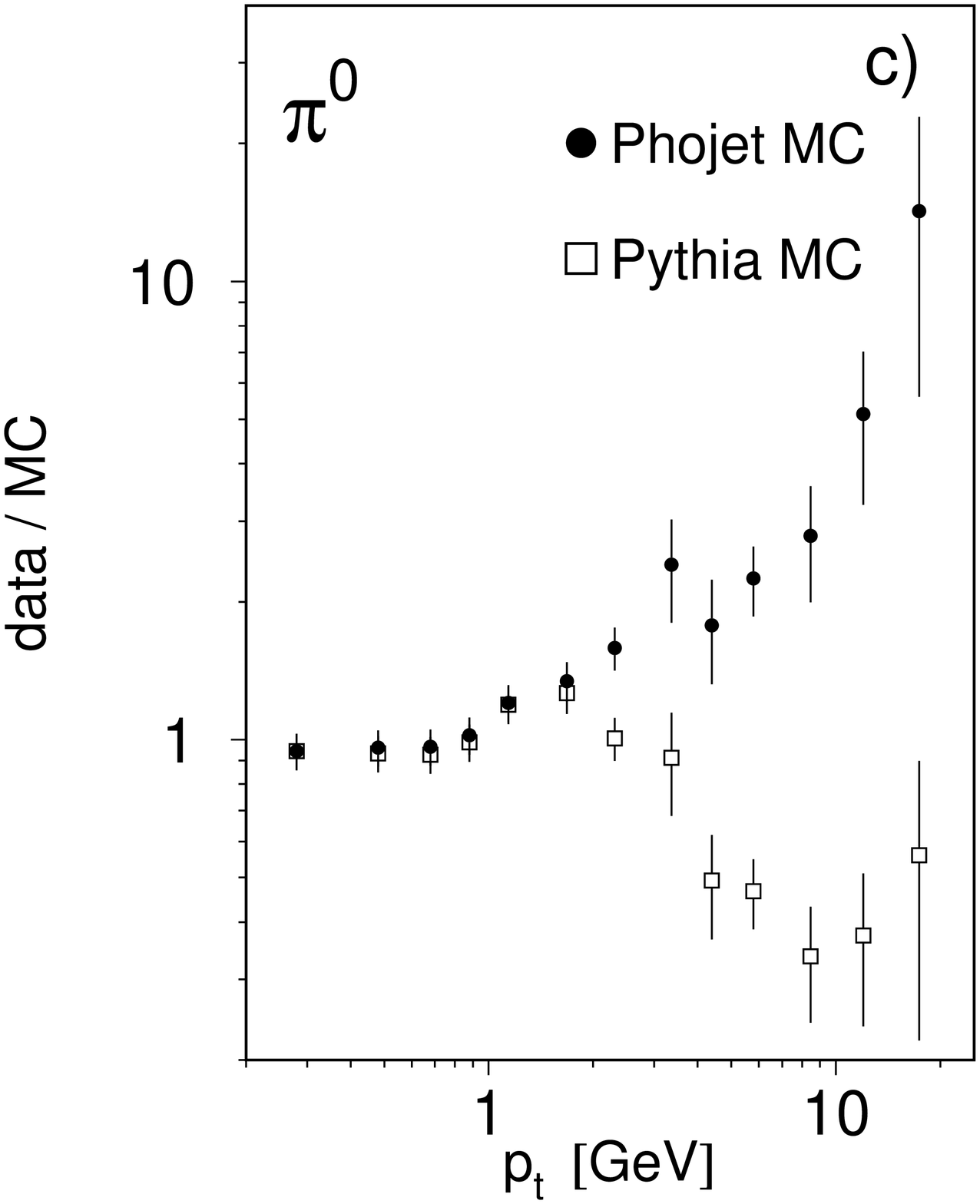}& 
    \includegraphics[width=0.45\figwidth]{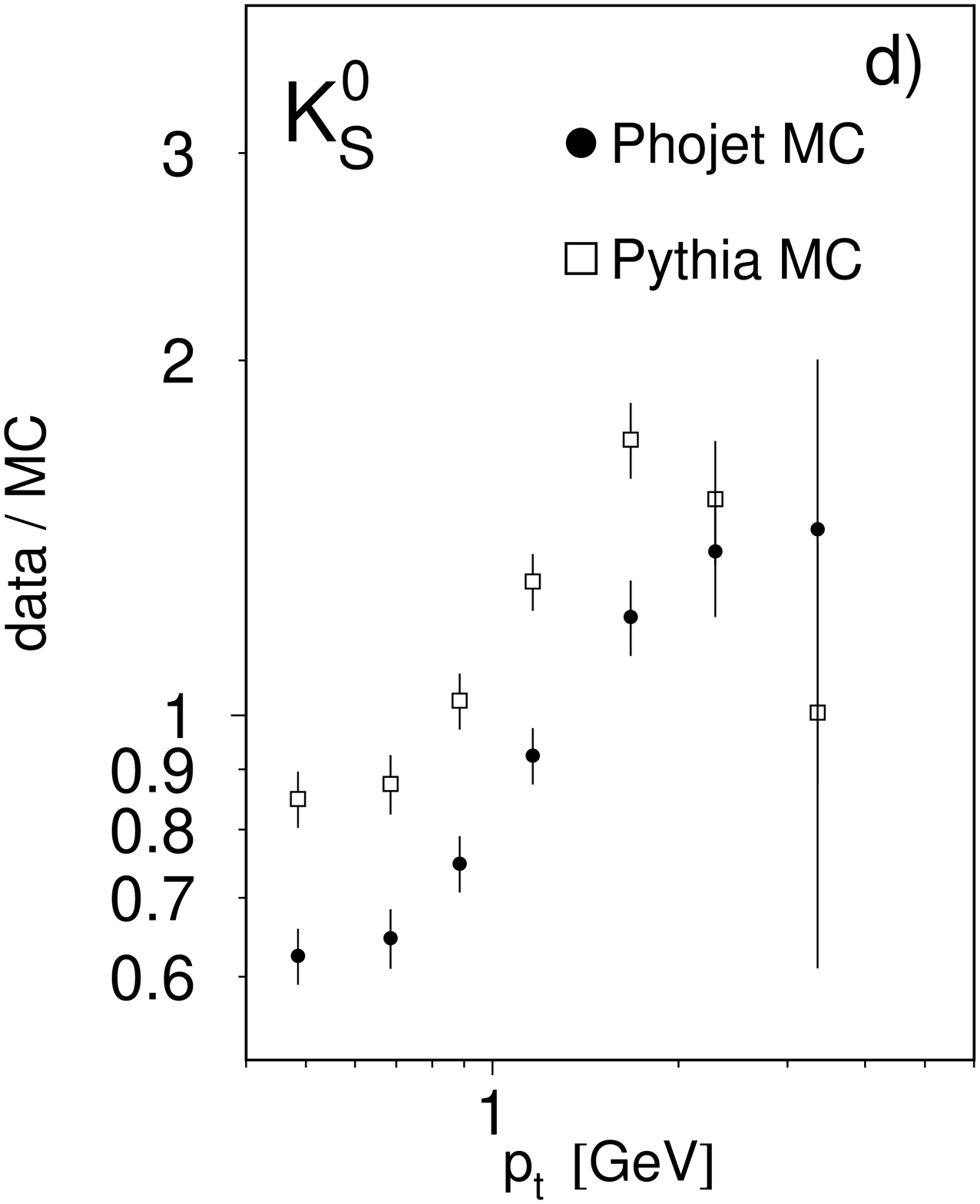}\\
 \end{tabular} 
  \caption{
  Inclusive differential cross section \dpt {} compared to Monte Carlo predictions
  and exponential and power law behaviour
  for: a) \pz {} production for $|\eta| <$0.5
  and b) \ks {} production for $|\eta| <$ 1.5.
  Ratio of the differential cross section \dpt {} to Monte Carlo predictions
  for: c) \pz {} production
  and d) \ks {} production.}
  \label{fig:ptMC2}
\end{figure}
\begin{sidewaysfigure}
  \begin{tabular}{cc}
    \includegraphics[width=0.45\linewidth]{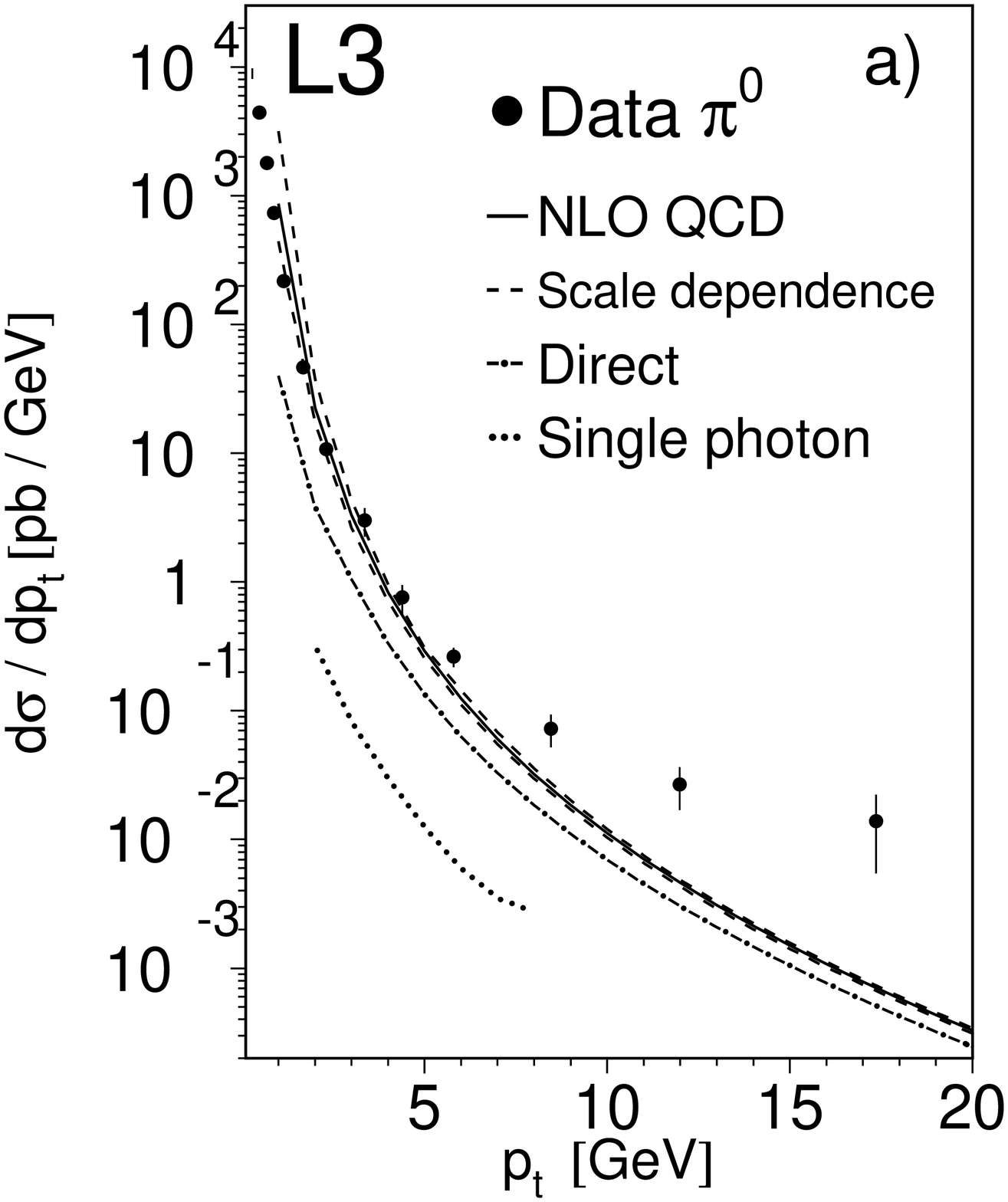}&
    \includegraphics[width=0.45\linewidth]{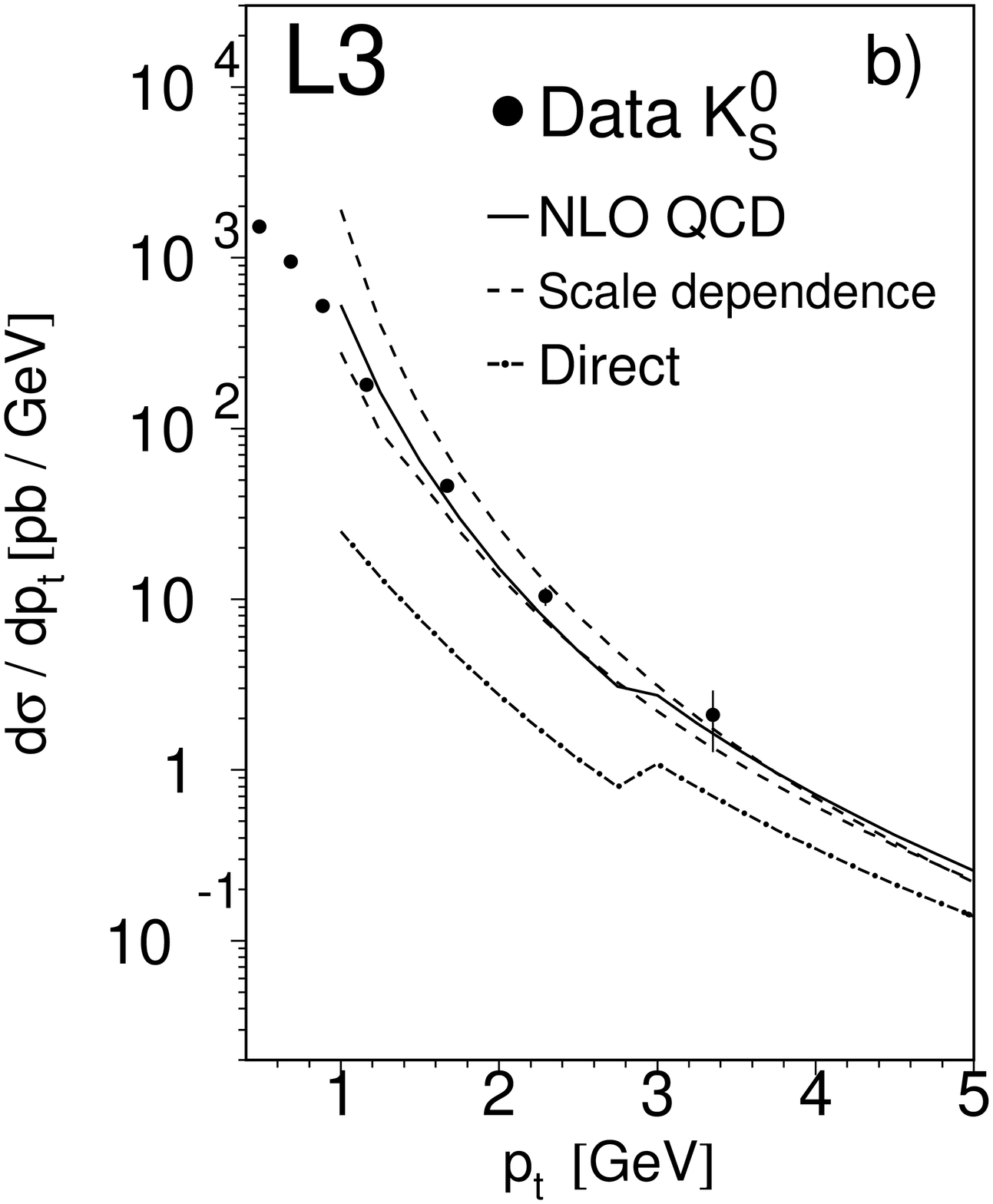}
  \end{tabular}
  \caption{Inclusive differential cross section \dpt {} 
  compared to NLO QCD predictions 
  for: a) \pz {} production for $|\eta| <$ 0.5 
  and b) \ks {} production for $|\eta| <$ 1.5.
  The NLO calculations are given for the QCD scale  equal to \pt {} (full line)
  and for the scales 0.5 \pt {} (upper dashed line) and 2 \pt {} (lower dashed line). 
  The contribution of the direct QED
  process is indicated as a dashed dotted line.
  For the \pz {} case the estimation of the single photon production~[16]
is indicated as a dotted line. The structure at 3\GeV{} in b) is
due to the charm threshold in the  fragmentation function [2,22].
}
  \label{fig:pt}
\end{sidewaysfigure}
\begin{sidewaysfigure}
  \begin{tabular}{cc}
  \includegraphics[width=0.45\linewidth]{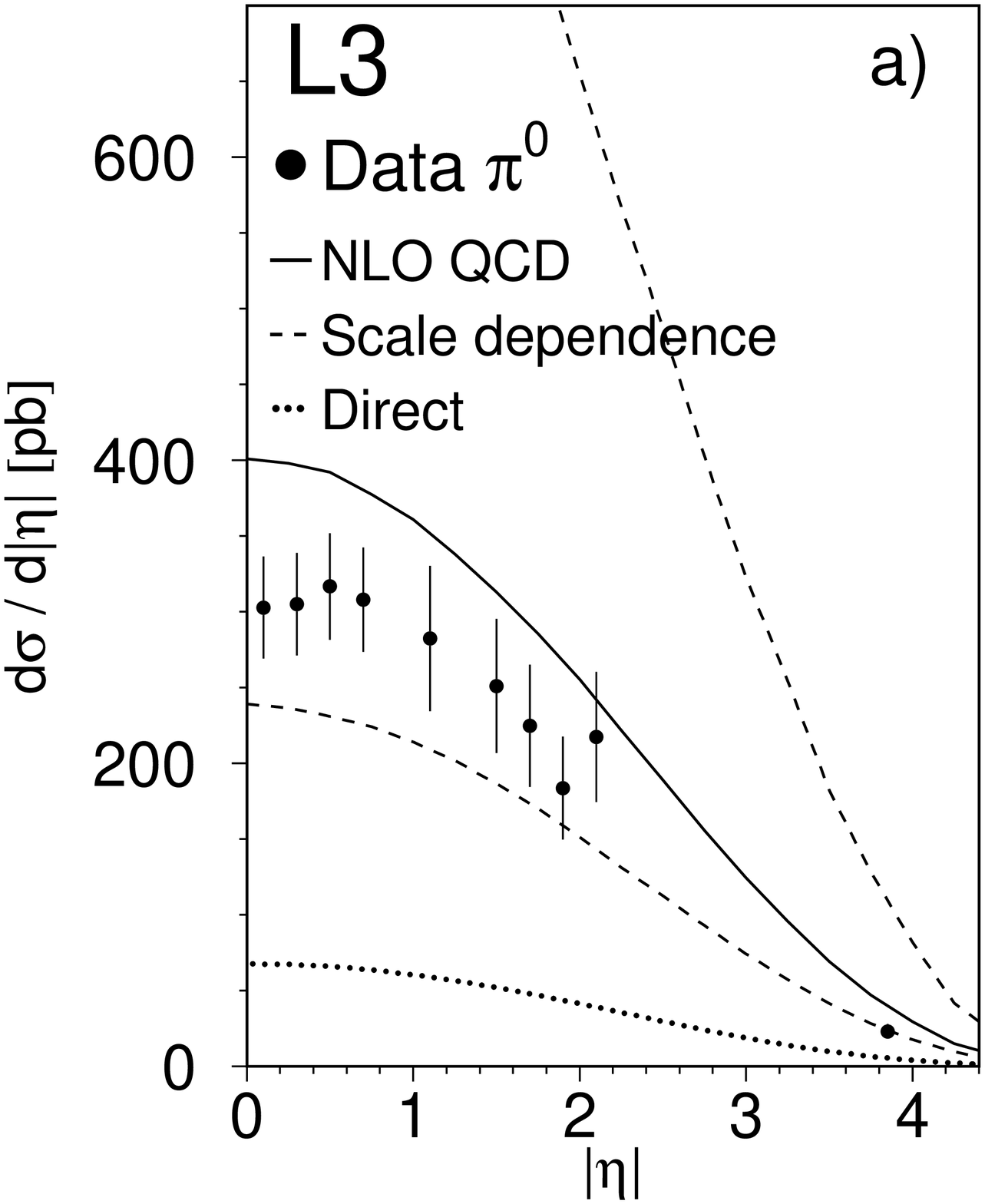}&
  \includegraphics[width=0.45\linewidth]{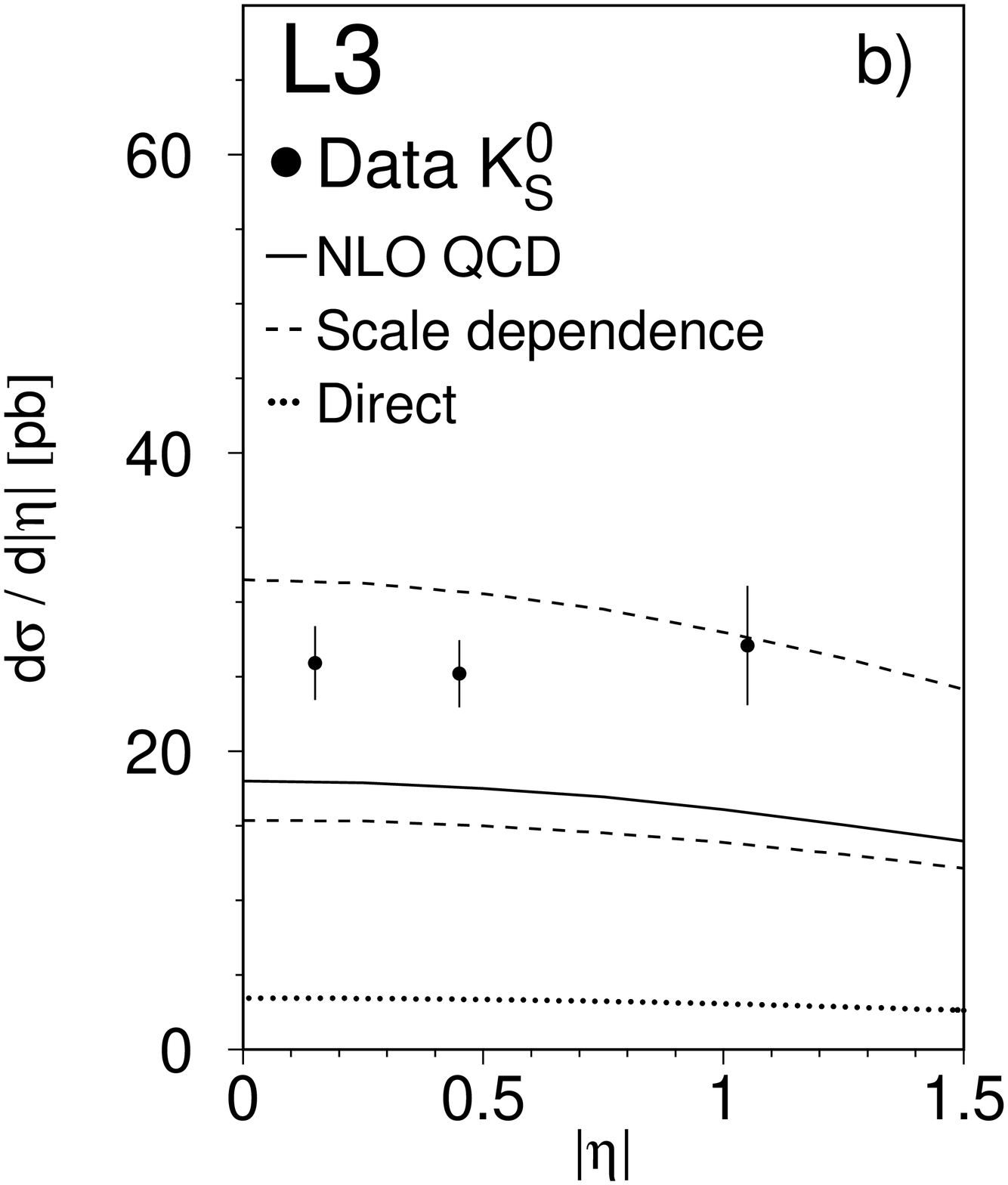}\\
  \end{tabular} 
  \caption{Inclusive differential cross sections \deta {} 
  compared to NLO QCD predictions
  for: a) \pz {} production for $\pt >$ 1 \GeV {}
  and b) \ks {} production for $\pt >$ 1.5 \GeV .
  The NLO calculations are given for the QCD scale  equal to \pt {} (full line)
  and for the scales 0.5 \pt {} (upper dashed line) and 2 \pt {} (lower dashed line). 
  The contribution of the direct QED
  process is indicated as a dotted line.}
  \label{fig:eta}
\end{sidewaysfigure}

Differential cross sections of \pz {} and \ks {} production
with respect to \pt {} and
$|\eta|$ are shown in Figures \ref{fig:ptMC2}, \ref{fig:pt} and \ref{fig:eta}. 

\par
The behaviour of \dpt {} 
in the range $0.2 < \pt < 1.5 \GeV ${} for \pz {} and
$0.6 < \pt < 1.5 \GeV $ for \ks {} is described by an exponential
 of the form $A e^{-\pt/\langle\pt\rangle}$
with a mean  value of \mbox{$\langle\pt\rangle \simeq 230$  \MeV} {} 
for the \pz {} and $\langle\pt\rangle \simeq 290$  \MeV {}
for the \ks .
This behaviour is characteristic  of hadrons produced by soft interactions  
and is similar to that obtained in hadron-hadron
and photon-hadron collisions \cite{perl}. 
Due to the direct $\Gg \ra \rm{q\bar{q}}$ process and to 
hard QCD interactions, two-photon
collisions exhibit a cross section 
higher than the expected exponential behaviour
at high \pt {} values. 
For  $\pt \ge 1.5 \GeV$, the  differential cross sections
are better represented 
by a power law function $A \pt^{-B}$.
The value of the power B is compatible with 4 for both
\pz {} and \ks .
In the framework of Reference \cite{brodsky},
this value is expected in the case of 
\mbox{$2 \rightarrow 2$} hard scattering processes at the parton level. 

\par
The differential cross sections are also compared to Monte Carlo predictions  
 in  Figure \ref{fig:ptMC2}. 
In the \pz {} case, the high \pt {} region is not reproduced by PYTHIA nor by PHOJET.
We verify that the shapes of the $|\eta|$ distributions of \pz {} and \ks {}
are well reproduced by both models.

\par
In Figures \ref{fig:pt}a and \ref{fig:pt}b
the data are compared
to  analytical NLO QCD predictions \cite{kniehl2}. For this calculation,
the flux of quasi-real  photons is obtained using
the Equivalent Photon Approximation, taking into account both     
transverse and  longitudinal virtual photons. 
The interacting particles can be photons or 
partons  from the $\gamma \ra \rm{q\bar{q}}$, which
evolves into quarks and gluons.
The NLO parton density functions of
Reference  \cite{aurenche} are used and
all elementary $2 \ra 2$ and $2\ra 3$ processes  are considered. 
New NLO fragmentation functions $(FF)$ \cite{FF}
are used  assuming that $FF(\pi^0) = (FF(\pi^+) + FF(\pi^-))/2$.
The renormalization, factorisation and fragmentation scales
are taken to be equal: $\mu=M=M_F=\xi \pt$ \cite{kniehl}.
The scale uncertainty in the NLO calculation is estimated by varying the value of $\xi$
from 0.5 to 2.0. 
The structure in the $\pt$ spectrum for the  \ks {} calculation is due to the charm 
threshold in the  fragmentation function~\cite{kniehl,kniehl3}.
The agreement with the data is satisfactory in the \ks {} case, but it is poor
for the \pz {} case in the high-\pt {} range. 
\par
The \deta {} differential cross sections, are also compared to QCD calculations
as shown in Figure \ref{fig:eta}. The shape of the data, and in particular
the measurement of the \pz {} production at $\langle |\eta | \rangle = 3.85$,
is reproduced by NLO QCD predictions.

%
%
\section*{Acknowledgements}

We would like to thank B. A. Kniehl, L. Gordon 
and M. Fontannaz for providing us with their NLO QCD calculations 
and R. Engel and T. Sj\"ostrand for useful discussions.

%
%

%
\newpage
\section*{Author List}
\typeout{   }     
\typeout{Using author list for paper 243 -- ? }
\typeout{$Modified: Jul 31 2001 by smele $}
\typeout{!!!!  This should only be used with document option a4p!!!!}
\typeout{   }
%
%
%
%
%
%

\newcount\tutecount  \tutecount=0
\def\tutenum#1{\global\advance\tutecount by 1 \xdef#1{\the\tutecount}}
\def\tute#1{$^{#1}$}
\tutenum\aachen            
\tutenum\nikhef            
\tutenum\mich              
\tutenum\lapp              
\tutenum\basel             
\tutenum\lsu               
\tutenum\beijing           
\tutenum\berlin            
\tutenum\bologna           
\tutenum\tata              
\tutenum\ne                
\tutenum\bucharest         
\tutenum\budapest          
\tutenum\mit               
\tutenum\panjab            
\tutenum\debrecen          
\tutenum\florence          
\tutenum\cern              
\tutenum\wl                
\tutenum\geneva            
\tutenum\hefei             
\tutenum\lausanne          
\tutenum\lyon              
\tutenum\madrid            
\tutenum\florida           
\tutenum\milan             
\tutenum\moscow            
\tutenum\naples            
\tutenum\cyprus            
\tutenum\nymegen           
\tutenum\caltech           
\tutenum\perugia           
\tutenum\peters            
\tutenum\cmu               
\tutenum\potenza           
\tutenum\prince            
\tutenum\riverside         
\tutenum\rome              
\tutenum\salerno           
\tutenum\ucsd              
\tutenum\sofia             
\tutenum\korea             
\tutenum\utrecht           
\tutenum\purdue            
\tutenum\psinst            
\tutenum\zeuthen           
\tutenum\eth               
\tutenum\hamburg           
\tutenum\taiwan            
\tutenum\tsinghua          

{
\parskip=0pt
\noindent
{\bf The L3 Collaboration:}
\ifx\selectfont\undefined
 \baselineskip=10.8pt
 \baselineskip\baselinestretch\baselineskip
 \normalbaselineskip\baselineskip
 \ixpt
\else
 \fontsize{9}{10.8pt}\selectfont
\fi
\medskip
\tolerance=10000
\hbadness=5000
\raggedright
\hsize=162truemm\hoffset=0mm
\def\r{\rlap,}
\noindent

P.Achard\r\tute\geneva\ 
O.Adriani\r\tute{\florence}\ 
M.Aguilar-Benitez\r\tute\madrid\ 
J.Alcaraz\r\tute{\madrid,\cern}\ 
G.Alemanni\r\tute\lausanne\
J.Allaby\r\tute\cern\
A.Aloisio\r\tute\naples\ 
M.G.Alviggi\r\tute\naples\
H.Anderhub\r\tute\eth\ 
V.P.Andreev\r\tute{\lsu,\peters}\
% ...
\newpage
\begin{list}{A}{\itemsep=0pt plus 0pt minus 0pt\parsep=0pt plus 0pt minus 0pt
                \topsep=0pt plus 0pt minus 0pt}
\item[\aachen]
 I. Physikalisches Institut, RWTH, D-52056 Aachen, FRG$^{\S}$\\
 III. Physikalisches Institut, RWTH, D-52056 Aachen, FRG$^{\S}$
\item[\nikhef] National Institute for High Energy Physics, NIKHEF, 
     and University of Amsterdam, NL-1009 DB Amsterdam, The Netherlands
\item[\mich] University of Michigan, Ann Arbor, MI 48109, USA
\item[\lapp] Laboratoire d'Annecy-le-Vieux de Physique des Particules, 
     LAPP,IN2P3-CNRS, BP 110, F-74941 Annecy-le-Vieux CEDEX, France
\item[\basel] Institute of Physics, University of Basel, CH-4056 Basel,
     Switzerland
\item[\lsu] Louisiana State University, Baton Rouge, LA 70803, USA
\item[\beijing] Institute of High Energy Physics, IHEP, 
  100039 Beijing, China$^{\triangle}$ 
\item[\berlin] Humboldt University, D-10099 Berlin, FRG$^{\S}$
\item[\bologna] University of Bologna and INFN-Sezione di Bologna, 
     I-40126 Bologna, Italy
\item[\tata] Tata Institute of Fundamental Research, Mumbai (Bombay) 400 005, India
\item[\ne] Northeastern University, Boston, MA 02115, USA
\item[\bucharest] Institute of Atomic Physics and University of Bucharest,
     R-76900 Bucharest, Romania
\item[\budapest] Central Research Institute for Physics of the 
     Hungarian Academy of Sciences, H-1525 Budapest 114, Hungary$^{\ddag}$
\item[\mit] Massachusetts Institute of Technology, Cambridge, MA 02139, USA
\item[\panjab] Panjab University, Chandigarh 160 014, India.
\item[\debrecen] KLTE-ATOMKI, H-4010 Debrecen, Hungary$^\P$
\item[\florence] INFN Sezione di Firenze and University of Florence, 
     I-50125 Florence, Italy
\item[\cern] European Laboratory for Particle Physics, CERN, 
     CH-1211 Geneva 23, Switzerland
\item[\wl] World Laboratory, FBLJA  Project, CH-1211 Geneva 23, Switzerland
\item[\geneva] University of Geneva, CH-1211 Geneva 4, Switzerland
\item[\hefei] Chinese University of Science and Technology, USTC,
      Hefei, Anhui 230 029, China$^{\triangle}$
\item[\lausanne] University of Lausanne, CH-1015 Lausanne, Switzerland
\item[\lyon] Institut de Physique Nucl\'eaire de Lyon, 
     IN2P3-CNRS,Universit\'e Claude Bernard, 
     F-69622 Villeurbanne, France
\item[\madrid] Centro de Investigaciones Energ{\'e}ticas, 
     Medioambientales y Tecnolog{\'\i}cas, CIEMAT, E-28040 Madrid,
     Spain${\flat}$ 
\item[\florida] Florida Institute of Technology, Melbourne, FL 32901, USA
\item[\milan] INFN-Sezione di Milano, I-20133 Milan, Italy
\item[\moscow] Institute of Theoretical and Experimental Physics, ITEP, 
     Moscow, Russia
\item[\naples] INFN-Sezione di Napoli and University of Naples, 
     I-80125 Naples, Italy
\item[\cyprus] Department of Physics, University of Cyprus,
     Nicosia, Cyprus
\item[\nymegen] University of Nijmegen and NIKHEF, 
     NL-6525 ED Nijmegen, The Netherlands
\item[\caltech] California Institute of Technology, Pasadena, CA 91125, USA
\item[\perugia] INFN-Sezione di Perugia and Universit\`a Degli 
     Studi di Perugia, I-06100 Perugia, Italy   
\item[\peters] Nuclear Physics Institute, St. Petersburg, Russia
\item[\cmu] Carnegie Mellon University, Pittsburgh, PA 15213, USA
\item[\potenza] INFN-Sezione di Napoli and University of Potenza, 
     I-85100 Potenza, Italy
\item[\prince] Princeton University, Princeton, NJ 08544, USA
\item[\riverside] University of Californa, Riverside, CA 92521, USA
\item[\rome] INFN-Sezione di Roma and University of Rome, ``La Sapienza",
     I-00185 Rome, Italy
\item[\salerno] University and INFN, Salerno, I-84100 Salerno, Italy
\item[\ucsd] University of California, San Diego, CA 92093, USA
\item[\sofia] Bulgarian Academy of Sciences, Central Lab.~of 
     Mechatronics and Instrumentation, BU-1113 Sofia, Bulgaria
\item[\korea]  The Center for High Energy Physics, 
     Kyungpook National University, 702-701 Taegu, Republic of Korea
\item[\utrecht] Utrecht University and NIKHEF, NL-3584 CB Utrecht, 
     The Netherlands
\item[\purdue] Purdue University, West Lafayette, IN 47907, USA
\item[\psinst] Paul Scherrer Institut, PSI, CH-5232 Villigen, Switzerland
\item[\zeuthen] DESY, D-15738 Zeuthen, 
     FRG
\item[\eth] Eidgen\"ossische Technische Hochschule, ETH Z\"urich,
     CH-8093 Z\"urich, Switzerland
\item[\hamburg] University of Hamburg, D-22761 Hamburg, FRG
\item[\taiwan] National Central University, Chung-Li, Taiwan, China
\item[\tsinghua] Department of Physics, National Tsing Hua University,
      Taiwan, China
\item[\S]  Supported by the German Bundesministerium 
        f\"ur Bildung, Wissenschaft, Forschung und Technologie
\item[\ddag] Supported by the Hungarian OTKA fund under contract
numbers T019181, F023259 and T024011.
\item[\P] Also supported by the Hungarian OTKA fund under contract
  number T026178.
\item[$\flat$] Supported also by the Comisi\'on Interministerial de Ciencia y 
        Tecnolog{\'\i}a.
\item[$\sharp$] Also supported by CONICET and Universidad Nacional de La Plata,
        CC 67, 1900 La Plata, Argentina.
\item[$\triangle$] Supported by the National Natural Science
  Foundation of China.
\end{list}
}
\vfill


%
%
\newpage

\begin{table}
  \begin{center}

    \begin{tabular}{|r@{$-$}l|c||r@{~$\pm$~}l|r@{~$\pm$~}r@{~$\pm$~}lc||r@{~$\pm$~}r@{~$\pm$~}lc}
    \hline
    \multicolumn{2}{|c|}{\pt}  & $\langle\pt\rangle$ &  \multicolumn{2}{|c|}{Efficiency}
    & \multicolumn{4}{|c||}{\dpt {} for $\Wgg > 5$ GeV}
    & \multicolumn{4}{|c|}{\dpt {}  for $\Wgg > 10$ GeV}\\
     \multicolumn{2}{|c|}{[GeV]}  & [GeV]& \multicolumn{2}{|c|}{[\%] } 
     & \multicolumn{4}{|c||}{ [pb/GeV] }
     & \multicolumn{4}{|c|}{ [pb/GeV] }\\
    \hline
    0.2&0.4 &  \phantom{0}0.28 & 12.9 & 1.2 & (89 & \phantom{0}1\phantom{.0} & \phantom{0}8) & $\times 10^2$    & (62 & 0.8            & \phantom{0}6) & \multicolumn{1}{c|}{$\times 10^2$}\\
    \hline
    0.4&0.6 &  \phantom{0}0.48 & 24.3 & 2.2 & (44 & \phantom{0}0.3           & \phantom{0}4) & $\times 10^2$    & (35 & 0.3            & \phantom{0}3) & \multicolumn{1}{c|}{$\times 10^2$}\\
    \hline
    0.6&0.8 &  \phantom{0}0.68 & 30.7 & 2.8 & (18 & \phantom{0}0.1           & \phantom{0}2) & $\times 10^2$    & (15 & 0.1            & \phantom{0}1) & \multicolumn{1}{c|}{$\times 10^2$}\\
    \hline
    0.8&1.0 &  \phantom{0}0.88 & 35.4 & 3.2 & (73 & \phantom{0}0.8           & \phantom{0}7) & $\times 10^1$    & (59 & 0.7            & \phantom{0}6) & \multicolumn{1}{c|}{$\times 10^1$}\\
    \hline
    1.0&1.5 &  \phantom{0}1.14 & 37.2 & 3.4 & (22 & \phantom{0}0.3           & \phantom{0}2) & $\times 10^1$    & (18 & 0.3            & \phantom{0}2) & \multicolumn{1}{c|}{$\times 10^1$}\\
    \hline
    1.5&2.0 &  \phantom{0}1.68 & 37.4 & 3.5 & (46 & \phantom{0}1\phantom{.0} & \phantom{0}4) &                  & (40 & 1\phantom{.0}  & \phantom{0}4) &  \multicolumn{1}{c|}{ }\\
    \hline
    2.0&3.0 &  \phantom{0}2.31 & 35.8 & 3.5 & (11 & \phantom{0}0.5           & \phantom{0}1) &                  & (95 & 5\phantom{.0}  &           11) &  \multicolumn{1}{c|}{$\times 10^{-1}$}\\
    \hline
    3.0&4.0 &  \phantom{0}3.36 & 23.5 & 4.1 & (30 & \phantom{0}6\phantom{.0} & \phantom{0}5) & $\times 10^{-1}$ & \multicolumn{4}{c}{}       \\
    \cline{1-9}
    4.0&5.0 &  \phantom{0}4.39 & 47.5 & 8.5 & (76 &           14\phantom{.0} & \phantom{0}1) & $\times 10^{-2}$ & \multicolumn{4}{c}{}       \\
    \cline{1-9}
    5.0&7.5 &  \phantom{0}5.79 & 26.7 & 3.0 & (26 & \phantom{0}4\phantom{.0} & \phantom{0}3) & $\times 10^{-2}$ & \multicolumn{4}{c}{}       \\
    \cline{1-9}
    7.5&10.0&  \phantom{0}8.46 & 26.4 & 3.7 & (73 &           18\phantom{.0} &           10) & $\times 10^{-3}$ & \multicolumn{4}{c}{}        \\
    \cline{1-9}
    10.0&15.0&           11.98 & 21.7 & 3.9 & (27 & \phantom{0}9\phantom{.0} & \phantom{0}5) & $\times 10^{-3}$ & \multicolumn{4}{c}{}        \\
    \cline{1-9} 
     15.0&20.0&          17.36 & 15.6 & 3.8 & (14 & \phantom{0}8\phantom{.0} & \phantom{0}4) & $\times 10^{-3}$  & \multicolumn{4}{c}{}       \\
    \cline{1-9}
    \end{tabular}

    \caption{The \pz {} overall efficiency and differential cross sections 
    as a function of 
    \pt {}  for $|\eta| <$ 0.5.
    For $\pt <$ 4 GeV, the \pz {} is only reconstructed from
    its decay into two photons. Above 5 GeV,
    the \pz {} is only detected as a single cluster. 
    In the $4-5$ GeV bin, both methods are used yielding a higher efficiency. 
    The first uncertainty on cross sections
    is statistical and the second one systematic.
    The cross sections are calculated for $\Wgg >$ 5 GeV and $\Wgg >$ 10 GeV
    and coincide for $\pt >$ 3 GeV.}
    \label{tab:pi0pt}

  \end{center}
\end{table}

\begin{table}
  \begin{center}

    \begin{tabular}{|c|c||c|r@{~$\pm$~}l|r@{~$\pm$~}r@{~$\pm$~}l|}
    \hline
    Detector & $|\eta|$  & Number of \pz & \multicolumn{2}{|c|}{Efficiency [\%]} 
    & \multicolumn{3}{|c|}{\deta {} [pb]}\\
    \hline
    & 0.0$-$0.2  & 8914 & \hspace{.12 cm} 35.6 & 4.0 & 303 & \phantom{0}8 & 33       \\
    \cline{2-8}
    Barrel & 0.2$-$0.4  & 9263 & 36.7 & 4.1 & 305 &\phantom{0} 8 & 33      \\
    \cline{2-8}
    & 0.4$-$0.6  & 8965 & 34.2 & 3.8 & 317 & \phantom{0}8 & 34          \\
    \cline{2-8}
    & 0.6$-$0.8  & 8094 & 31.7 & 3.6 & 308 & \phantom{0}8 & 33     \\
    \hline
    & 0.8$-$1.4  & 8688 & 12.4 & 2.1 & 282 & 10 & 47     \\
    \cline{2-8}
    & 1.4$-$1.6  & 3443 & 16.6 & 2.9 & 251 & 15 & 42      \\
    \cline{2-8}
    Endcap & 1.6$-$1.8  & 3050 & 16.4 & 3.0 & 225 & 15 & 37  \\
    \cline{2-8}
    & 1.8$-$2.0  & 2313 & 15.2 & 2.8 & 184 & 15 & 31    \\
    \cline{2-8}
    & 2.0$-$2.2  & 2294 & 12.7 & 2.5 & 217 & 23 & 36    \\
    \hline
    Small-angle & 3.4$-$4.3  & 1410 & 16.4 & 3.5 &  23 & 2 & \phantom{0}5     \\
    \hline
    \end{tabular}

    \caption{The number of reconstructed \pz,
    overall efficiency and differential cross section 
    as a function of pseudorapidity
    for $\pt > 1$ GeV and $\Wgg > 5$ GeV. 
    The first uncertainty on the cross section 
    is statistical and the second one systematic.}
    \label{tab:pi0eta}

  \end{center}
\end{table}

\newpage

\begin{table}
  \begin{center}

    \begin{tabular}{|c|c||r@{~$\pm$~}l|r@{~$\pm$~}r@{~$\pm$~}l||r@{~$\pm$~}r@{~$\pm$~}l|}
    \hline
    \pt& $\langle\pt\rangle$   &  \multicolumn{2}{|c|}{Efficiency}
    & \multicolumn{3}{|c||}{\dpt {} for $\Wgg > 5$ GeV}
    & \multicolumn{3}{|c|}{\dpt {}  for $\Wgg > 10$ GeV}\\
     \multicolumn{1}{|c|}{[GeV]} & [GeV] & \multicolumn{2}{|c|}{[\%] }   
     & \multicolumn{3}{|c||}{ [pb/GeV] }
     & \multicolumn{3}{|c|}{ [pb/GeV] }\\
    \hline
    0.4$-$0.6  & 0.49 & 13.6  & 0.1 & \hspace{.1 cm} 1522\phantom{.0} & 20\phantom{.0} &           81   & \hspace{.2 cm} 1262 & 21\phantom{.0} & 65         \\
    \hline
    0.6$-$0.8  & 0.69 & 17.8  & 0.2 &                947\phantom{.0}  & 14\phantom{.0} &           53   &  786 &           16\phantom{.0} & 40     \\
    \hline
    0.8$-$1.0  & 0.89 & 20.0  & 0.3 &                521\phantom{.0}  & 11\phantom{.0} &           27   &  428 &           12\phantom{.0} & 22      \\
    \hline
    1.0$-$1.5  & 1.16 & 22.2  & 0.4 &                180\phantom{.0}  & 5\phantom{.0}  & \phantom{0}9   &  151 & \phantom{0}4\phantom{.0} & \phantom{0}8       \\
    \hline
    1.5$-$2.0  & 1.67 & 22.4  & 1.0 &                46\phantom{.0}   & 3\phantom{.0}  & \phantom{0}3   &  42  & \phantom{0}2\phantom{.0} & \phantom{0}2      \\
    \hline
    2.0$-$3.0  & 2.29 & 16.1  & 1.3 &                10\phantom{.0}   & 1\phantom{.0}  & \phantom{0}0.6 & 11  & \phantom{0}1\phantom{.0}  & \phantom{0}0.6    \\
    \hline
    3.0$-$4.0  & 3.35 & 14.8  & 4.8 &                2.1              & 0.8            & \phantom{0}0.2 & 1   & \phantom{0}0.5            & \phantom{0}0.3    \\
    \hline
    \end{tabular}

    \caption{The \ks {} overall efficiency and differential 
    cross sections as a function of \pt {}
    for $|\eta| < 1.5$.
    The first uncertainty on cross section
    is statistical and the second one systematic.    
    The cross section is calculated for $\Wgg >$ 5 GeV and $\Wgg >$ 10 GeV.}
    \label{tab:k0spt}

  \end{center}
\end{table}

\begin{table}
  \begin{center}

    \begin{tabular}{|c||c|r@{~$\pm$~}l|r@{~$\pm$~}l@{~$\pm$~}l|}
    \hline
    $|\eta|$  &  Number of \ks &  \multicolumn{2}{|c|}{Efficiency [\%]} 
    & \multicolumn{3}{|c|}{\deta [pb]}\\
    \hline
    0.0$-$0.3   & 744 & \hspace{.15 cm} 23.3  & 0.8 & 25.9 & 2.0  & 1.3      \\
    \hline
    0.3$-$0.6   & 759 & 24.5  & 0.8 & 25.1 & 1.9  & 1.2          \\
    \hline
    0.6$-$1.5   & 1473 & 14.7  & 1.2 & 27.1 & 2.4  & 3.2     \\
    \hline
    \end{tabular}

    \caption{The number of reconstructed \ks ,  overall efficiency
    and differential cross section as a function of pseudorapidity
    for $\pt > 1.5$ GeV and $\Wgg > 5$ GeV.
    The first uncertainty on the cross section
    is statistical and the second one systematic.}
    \label{tab:k0seta}

  \end{center}
\end{table}

\newpage

\end{document}